\documentclass[journal]{IEEEtran}

\usepackage{graphicx}
\usepackage{dcolumn}
\usepackage{bm}

\usepackage[dvips]{color}

\begin{document}
\title{An FET-Based Unit Cell for an Active Magnetic Metamaterial}

\author{Lukas~Jelinek
        and~Jan~Machac
\thanks{L.~Jelinek and J.~Machac are with the Department of Electromagnetic Field, Czech Technical University in Prague, 166 27-Prague, Czech Republic, e-mail: l\textunderscore jelinek@us.es.}
\thanks{Manuscript received June XXX, 2011; revised June XXX, 2011.}}

\markboth{XXX,~Vol.~XXX, No.~XXX, XXX~XXX}%
{L. Jelinek \MakeLowercase{\textit{et al.}}: XXX Title XXX}

\maketitle

\begin{abstract}A particle that can be used to create an active magnetic metamaterial has been designed using an FET transistor loaded in its gate by a conducting ring and in its source by a parallel resonance circuit. The design procedure is discussed and the working principle is experimentally demonstrated in the RF range. 
\end{abstract}

\begin{IEEEkeywords}
Artificial materials, FET circuits, Negative resistance devices, Polarizability, Oscillators
\end{IEEEkeywords}

\section{Introduction}

\IEEEPARstart{I}{n} the present state of the art, artificially made lattices of conducting rings terminated by a properly chosen impedance are considered a standard way of implementing magnetic metamaterials, i.e. artificial media with negative permeability, \cite{Marques, Tretyakov}. Materials made of conducting rings are also behind the practical realization of super-resolution lenses \cite{Freire-2008c} and of cloaking devices \cite{Schurig-2006}. Unfortunately, it is well known \cite{Marques-2004a,Schurig-2006} that the intrinsic losses of the rings and their loads are responsible for great degradation of lens and cloaking properties in comparison with their lossless theoretical proposals. Passive reduction of losses by a proper choice of their geometry is very limited \cite{Cummer-2008}, and thus active elements seem to be the only way out of this problem. The first proposal for using active elements in the ring metamaterial was published in 2001 \cite{Tretyakov-2001}, and used a Negative Impedance Converter (NIC) as a load. The NIC in \cite{Tretyakov-2001} was designed using an ideal operational amplifier, a component that works in a realistic implementation (with current technology) only up to the low RF range. A much simpler design of NIC has recently been used in \cite{Azab-2008, Hrabar-2010, Jin-2010}. A two-transistor implementation that offers a much wider frequency range of operation has been used in these works. In addition to loading rings with NIC, other authors have used two-port amplifiers to create an active magnetic metamaterial \cite{Popa-2007,Yuan-2009}. These designs however suffer from the need for input and output coils, which enlarge the unit cell and are necessarily coupled. In addition, the necessary two-port monolitic amplifier and phase shifter are difficult to make at high frequencies. 

The design presented in this paper uses a one-port approach, i.e. it uses a ring loaded by a one-port device offering at a given frequency a negative real part of impedance. Three main principles can be used to create this negative resistance. The first principle involves the Gunn diode \cite{Gunn-1963}, an element that is widely used for microwave oscillators up to sub-THz range. Unfortunately, the Gunn diode needs considerable cooling, as it works with a high DC bias, and the bulky nature of the sink preclude its use for metamaterial design. The second principle uses the above-mentioned NIC, which is a two-port device that images the loading impedance $Z$ on one port into an impedance $-Z$ on the other port \cite{Lilvill-1953}. A major review of possible NIC designs has been published in \cite{Sussman-1994, Sussman-1998}. It is shown that at least two transistors are needed for NIC design. The third principle is the use of a single transistor loaded in its source by a proper impedance, which makes it conditionally unstable. This is a common way of making HF oscillators, and it is the principle that we employ in our design. It offers the smallest size, the simplest assembly and even higher operating frequency than NIC. 

\section{Design approach}

Our design (see Fig. 1) is mostly inspired by \cite{Chung-2003}, where the structure was used for an active antenna array. As can be seen, the circuit is based on a JFET transistor connected in a common source mode, which is loaded at its gate by a ring and by a tuning impedance, and which has a parallel resonance circuit connected to its source. On the transistor gate, the negative resistance will appear when operating above the resonance frequency (the capacitive loading of the source) and will disappear at frequencies high enough to make a low impedance connection between gate and source (through the internal gate-source capacitance). The negative resistance is thus limited to a given frequency band. This is important in order to preclude self-maintained oscillations of the transistor, as will be discussed later. The tuning impedance in the gate serves to compensate the total impedance of the ring circuit in order to obtain the strongest possible response.
\begin{figure}
\centering
\includegraphics[width=0.7\columnwidth]{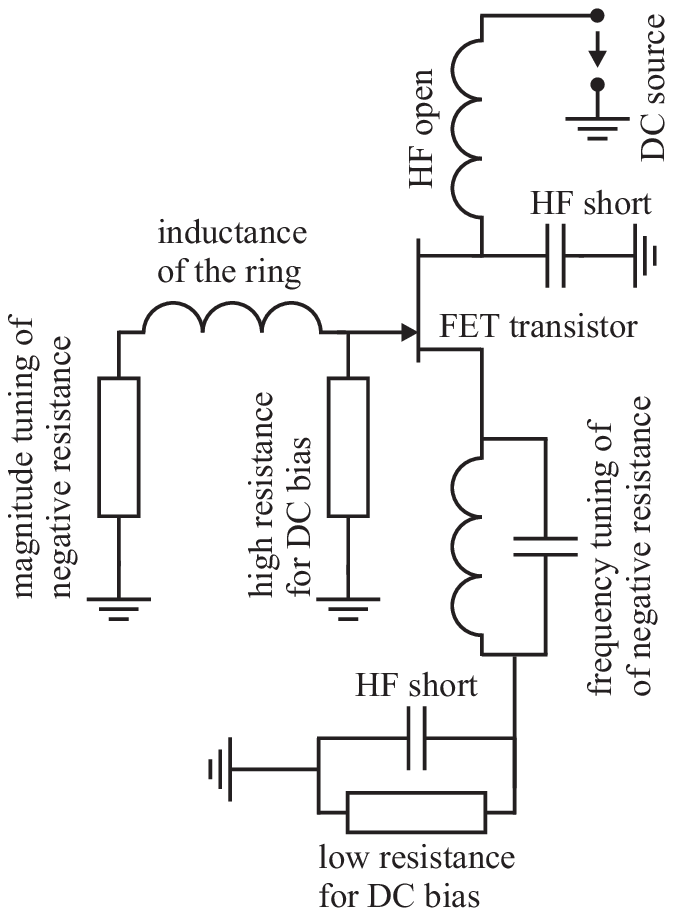}
\caption{\label{Fig1} Scheme of a ring loaded by a negative resistance transistor circuit. A JFET transistor is used, however any other FET transistor could be used.}
\end{figure}
The scenario from the point of view of the ring is depicted in Fig. 2. If properly tuned, the total impedance of the ring (the self inductance plus the load) should be small in absolute value to obtain a strong response and should have a negative real part in order to obtain a gain.
\begin{figure}
\centering
\includegraphics[width=0.8\columnwidth]{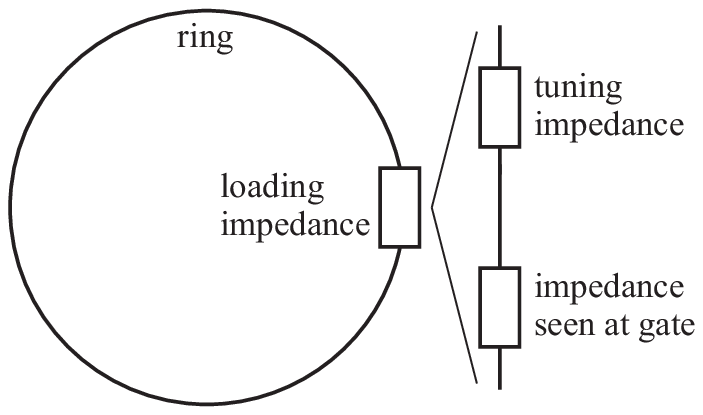}
\caption{\label{Fig2} Situation as seen by the ring.}
\end{figure}
The concept of Fig. 2 allows us to write the magnetic polarisability of the proposed particle as \cite{Tretyakov}
\begin{equation}
\alpha  = { -  \frac{{\rm{j}}\omega A^2 }{Z} } = { - \frac{{\rm{j}}\omega A^2 }{{\rm{j}}\omega L_{{\rm{ring}}}  + Z_{{\rm{tune}}}  + Z_{{\rm{gate}}} }}
\end{equation}
with $A$ as the area of the loop. For our purposes, the polarisability will be further normalized as
\begin{equation}
\alpha _{\rm{n}}  = \frac{{\mu _0 \alpha }}{V} =  - {\rm{j}}\frac{{2\pi ^3 N^2 \left( {r_0 /\lambda _0 } \right)}}{{\left( {Z/Z_0 } \right)\left( {V/r_0^3 } \right)}}
\end{equation}
where $V$ is the volume occupied by the particle, $r_0$ is the mean radius of the ring, which is assumed to be made of $N$ turns, $\lambda _0$ is a free space wavelength at the operating frequency, and $Z_0$ is the free space impedance. Using (2) a rough estimate of the permeability of a cubic lattice of such particles can be written as $\mu _{\rm{r}}  \approx 1 + \alpha _{\rm{n}}$.

\subsection{Frequency tuning and stability}

In the frequency range of intended operation, the impedance seen at the gate of the transistor has a capacitive imaginary part. From that point of view, the situation is identical to an ordinary Split Ring Resonator (SRR), where an inductive loop is loaded by a capacitance. However, the presented circuit, unlike in SRR, also imposes some positive or negative (depending on frequency) real part of the impedance ${\mathop{\rm Re}\nolimits} \left( Z  \right)$. Denoting $\omega _0$ as the resonance frequency of the loop circuit (see Fig. 2), the full list of possible responses is as follows
\begin{eqnarray}
\begin{array}{l}
 \omega  < \omega _0  \Rightarrow {\mathop{\rm Re}\nolimits} \left( \alpha  \right) > 0\;;\;\left| {\begin{array}{*{20}c}
   {{\mathop{\rm Im}\nolimits} \left( \alpha  \right) < 0 \;{\rm{for}} \;{\mathop{\rm Re}\nolimits} \left( Z  \right)  > 0}  \\
   {{\mathop{\rm Im}\nolimits} \left( \alpha  \right) > 0 \;{\rm{for}} \;{\mathop{\rm Re}\nolimits} \left( Z  \right) < 0}  \\
\end{array}} \right. \\ 
  \\ 
 \omega  = \omega _0  \Rightarrow {\mathop{\rm Re}\nolimits} \left( \alpha  \right) = 0\;;\;\left| {\begin{array}{*{20}c}
   {{\mathop{\rm Im}\nolimits} \left( \alpha  \right) < 0 \;{\rm{for}} \;{\mathop{\rm Re}\nolimits} \left( Z  \right) > 0}  \\
   {{\rm{unstable}} \;{\rm{for}} \;{\mathop{\rm Re}\nolimits} \left( Z  \right) < 0}  \\
\end{array}} \right. \\ 
  \\ 
 \omega  > \omega _0  \Rightarrow {\mathop{\rm Re}\nolimits} \left( \alpha  \right) < 0\;;\;\left| {\begin{array}{*{20}c}
   {{\mathop{\rm Im}\nolimits} \left( \alpha  \right) < 0 \;{\rm{for}} \;{\mathop{\rm Re}\nolimits} \left( Z  \right) > 0}  \\
   {{\mathop{\rm Im}\nolimits} \left( \alpha  \right) > 0 \;{\rm{for}} \;{\mathop{\rm Re}\nolimits} \left( Z  \right) < 0}  \\
\end{array}} \right. \\ 
 \end{array}
\end{eqnarray}
For our purposes (a magnetic metamaterial with negative real part of permeability), we are interested in the case of $\omega  > \omega _0$. However, care must be taken so that at $\omega  = \omega _0$ the real part of the impedance seen by the loop is positive, otherwise the circuit will be unstable and the particle will become a generator.

\section{Fabrication and measurement}

In order to prove the working principle of the proposed active ring, the frequency neighborhood of 100 MHz has been used, as it allows for easy hand-made implementation and it is close to the frequency used in common MRI machines - one of the important fields of metamaterial applications \cite{Wiltshire-2001,Freire-2008c}. A prototype was built using a JFET J310 transistor, which is commonly used in VHF amplifiers.  The parameters of the parallel resonant circuit in the source of the transistor were set to $f_{\rm{r}}  \approx 60\;{\rm{MHz}}$, $\sqrt {L/C} \approx 30\;\Omega$ and $R \approx 50\;\Omega$. After setting the bias so that the drain-source DC current was $I_{\rm{DS}} \approx 12\;{\rm{mA}}$, the impedance between the transistor gate and the common node was measured through the connected SMA connector on the vector network analyzer (VNA), see Fig. 3. The negative real part of the impedance can be appreciated in the frequency interval $75\;{\rm{MHz}} - 150\;{\rm{MHz}}$ together with the capacitive imaginary part. 
\begin{figure}
\centering
\includegraphics[width=0.95\columnwidth]{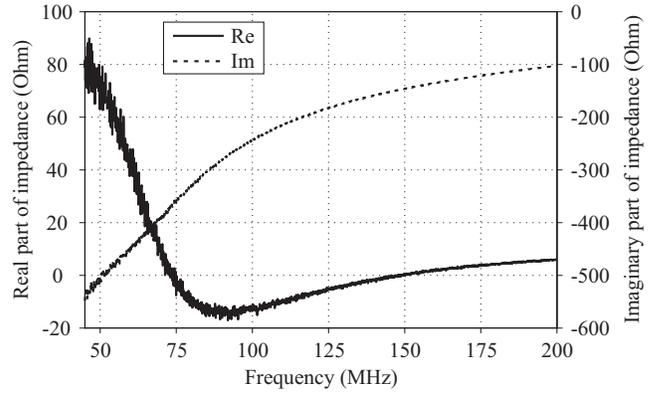}
\caption{\label{Fig3} Input impedance of the prototype at the transistor gate.}
\end{figure}
According to Sec. IIa, the imaginary part of the gate input impedance needs to be compensated just below $f \approx 75\;{\rm{MHz}}$, i.e. at a frequency just before the zero crossing of the real part. Only in this case will we get stable and strong magnetic polarisability with a negative real part and a positive imaginary part. It is also worth noting that in order to obtain low frequency dispersion and thus to make the compensation easier, the parallel resonant circuit in the source of the transistor is highly damped.

In this particular case, the compensation was made by the self inductance of the connected ring, which was actually made by 4 turns of wire with diameter $d = 0.2\;{\rm{mm}}$. The diameter of the loop was $D \approx 25\;{\rm{mm}}$. The fine frequency tuning was done by a slight change in the separation of the coil turns. The tuning impedance of Fig. 1 was not needed in this case. A photograph of the prototype is shown in Fig. 4.
\begin{figure}
\centering
\includegraphics[width=0.95\columnwidth]{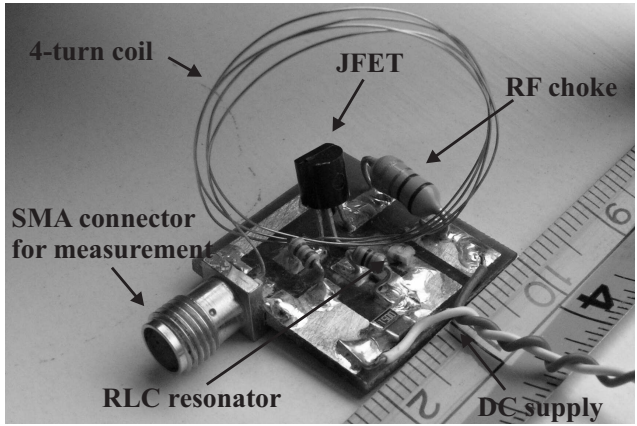}
\caption{\label{Fig4} Photograph of a realistic implementation of the ring loaded by a negative resistance circuit. The ring is actually a 4-turn coil made of thin wire.}
\end{figure}

With the compensating coil connected to the transistor gate, the impedance as seen by the coil induced voltage was once more measured on VNA via the SMA connector. The measured impedance was then substituted into (2) and the result is plotted in Fig. 5. The negative real part and the positive imaginary part of the normalized polarisability can be clearly observed in the figure in the frequency range of $73\;{\rm{MHz}} - 150\;{\rm{MHz}}$. The polarisability values shows that the real part of the permeability of a cubic lattice of such particles can easily be cast below zero. Note also that in our case the electrical size of the particle is rather small ($r_0 / \lambda_0 \approx  1/320$), and that according to (2) the response will be much stronger for particles of common metamaterial sizes ($r_0 / \lambda_0 \approx  1/20$).
\begin{figure}
\centering
\includegraphics[width=0.95\columnwidth]{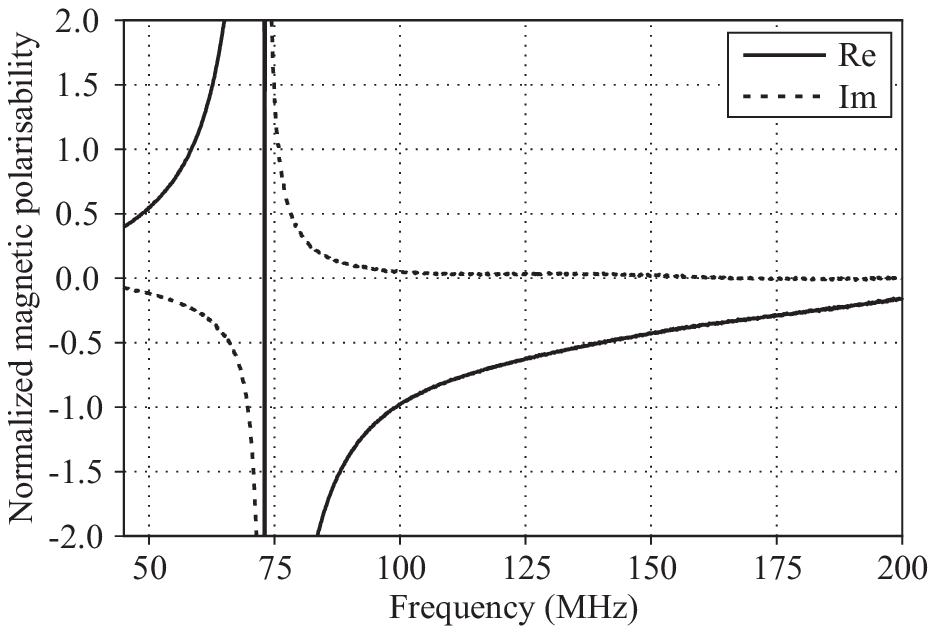}
\caption{\label{Fig5} Frequency dependence of the normalized magnetic polarisability of the particle. The normalization volume was taken as a brick with the external dimensions of the prototype, which is $20\;{\rm{mm}}\;{\rm{x}}\;20\;{\rm{mm}}\;{\rm{x}}\;30\;{\rm{mm}}$.}
\end{figure}

Until now the particle has been measured by directly connected VNA and thus not in the free space configuration in which it is intended to work in a realistic metamaterial. To this point the SMA connector has been substituted by a short. The VNA has then been coupled to the particle via a mutual inductance between the active ring and a loop of similar radius connected to the VNA port. The measured amplitude of the reflection coefficient is shown in Fig. 6, from where the amplification peak in the vicinity of $78\;{\rm{MHz}}$ can be seen, proving the principle proposed in this paper. In the same setup, the coupling loop has also been connected to a spectrum analyzer. It has been checked that the particle does not self generate any oscillations in the range of $9\;{\rm{kHz}} - 2\;{\rm{GHz}}$ and that what we have at hand is really an active metamaterial element and not a microwave generator.

\begin{figure}
\centering
\includegraphics[width=0.95\columnwidth]{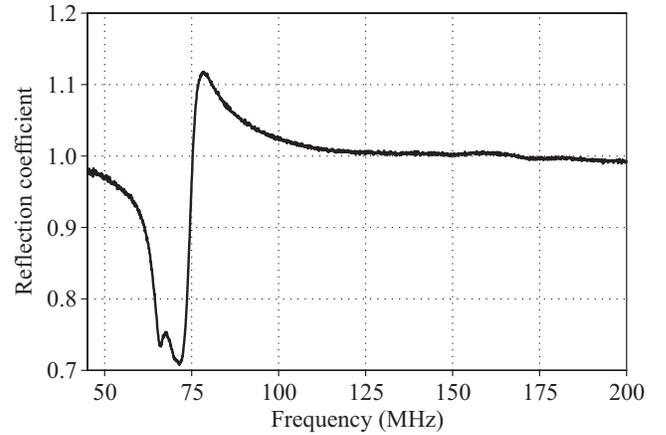}
\caption{\label{Fig6} The amplitude of the reflection coefficient measured on the loop magnetically coupled to the active ring.}
\end{figure}

\section{Conclusion}

An active particle using a ring loaded by an FET transistor circuit and offering magnetic polarisability with a negative real part and  a positive imaginary part has been proposed, designed and measured. The measurements show that the particle is capable of being used in negative permeability metamaterial systems, where it can remove losses or even add a gain. The use of the proposed particle in magnetic lenses and magnetic meta-surfaces is envisaged and should be studied in future.

\section*{Acknowledgment}

This work has been supported by the Czech Grant Agency (project No. 102/09/0314) and by the Czech Technical University in Prague (project No. SGS10/271/OHK3/3T/13). We would like to thank P. Cerny and M. Prihoda from the Department of Electromagnetic Field at Czech Technical University in Prague for their help with the measurements.

\bibliographystyle{IEEEtran}
\bibliography{IEEEabrv,References}

\end{document}